\documentclass[twocolumn,amsmath,amssymb,prl]{revtex4}
\usepackage{latexsym}
\usepackage{graphicx}
\usepackage{psfig}
\usepackage{color}
\usepackage[colorinlistoftodos]{todonotes}
\usepackage{verbatim}

\usepackage[letterpaper, margin=1in]{geometry}
\begin{document}

\title{Experimental Measurements of Effective Mass in Near Surface InAs Quantum Wells}
\author{Joseph~Yuan$^{1}$}
\author{Mehdi~Hatefipour$^{1}$}
\author{Brenden~A.~Magill$^{2}$}
\author{William~Mayer$^{1}$}
\author{Matthieu~C.~Dartiailh$^{1}$}
\author{Kasra~Sardashti$^{1}$}
\author{Kaushini~S.~Wickramasinghe$^{1}$}
\author{Giti~A.~Khodaparast$^{2}$}
\author{Yasuhiro~H.~Matsuda$^{3}$}
\author{Yoshimitsu~Kohama$^{3}$}
\author{Zhuo~Yang$^{3}$}
\author{Sunil~Thapa$^{4}$}
\author{Christopher~J.~Stanton$^{4}$}
\author{Javad~Shabani$^{1}$}

\affiliation{$^{1}$Center for Quantum Phenomena, Department of Physics, New York University, NY 10003, USA
\\
$^{2}$ Department of Physics, Virginia Tech, Blacksburg, VA 24061,USA
\\
$^{3}$ Institute for Solid State Physics, University of Tokyo, 5-1-5 Kashiwanoha, Kashiwa, Chiba 277-8581, Japan
\\
$^{4}$ Department of Physics, University of Florida, Gainesville, Florida 32611, USA
}

\date{\today}

\begin{abstract}
Near surface indium arsenide quantum wells have recently attracted a great deal of interest since they can be interfaced epitaxially with superconducting films and have proven to be a robust platform for exploring mesoscopic and topological superconductivity. In this work, we present magnetotransport properties of two-dimensional electron gases confined to an indium arsenide quantum well near the surface. The electron mass extracted from the envelope of the Shubnikov-de Haas oscillations shows an average effective mass $m^{*}$ = 0.04 at low magnetic field. Complementary to our magnetotransport study, we employed cyclotron resonance measurements and extracted the electron effective mass in the ultra high magnetic field regime. Both regimes can be understood by considering a model that includes non-parabolicity of the indium arsenide conduction bands. 
\end{abstract}


\maketitle
\subsection{I. Introduction}
Wafer-scale methods for the epitaxial growth of thin films of Aluminum (Al) on Indium Arsenide (InAs) heterostructures have recently been developed which yield uniform and atomically flat interfaces \cite{Kaushini2018,Shabani2016, JoonSue, Pauka2019}.  Josephson junctions fabricated on these materials yield a gate-controllable supercurrent with highly transparent contacts between the Al top layer and an InAs quantum well (QW) directly below the surface \cite{MortenPRA2017,Henri17,Billy2019,Lee2019,Fabrizio17}. Tuning of the semiconductor properties will affect supercurrent and other superconducting properties due to the wavefunction overlap at the  epitaxial interface.  Josephson junctions made out of Al-InAs have been used for tunable superconducting qubits, the so-called ``gatemon'' where the Josephson energy can be tuned in-situ with an applied electric field \cite{Larsen15, Casparis2018}. Furthermore, since InAs has large spin-orbit coupling, they can host topological superconductivity and Majorana bound states \cite{2019Mayer_Mat, mayer2019anom, Ren2019, FornieriNature2019}. The key feature in these structures is that the two-dimensional electron gases (2DEG) is confined near the surface, in close proximity to the superconductor.  While the epitaxial interface creates high contact transparency, it is expected that electron mobility of the 2DEG deteriorates due to increased rates of surface scattering as compared to isolated 2DEGs buried beneath the surface \cite{Kaushini2018,hatke,Shayegan2017}.  The myriad of possible applications with this platform implores a deeper study of the characteristics and material properties for near surface InAs QWs.  In this work, the transport experiments investigate the isolated semiconductor with the superconducting layer removed and the optical measurements are conducted on the semiconductor samples which did not have a superconducting layer to begin with.

\begin{figure*}[t]
\centering
\includegraphics[width=\textwidth]{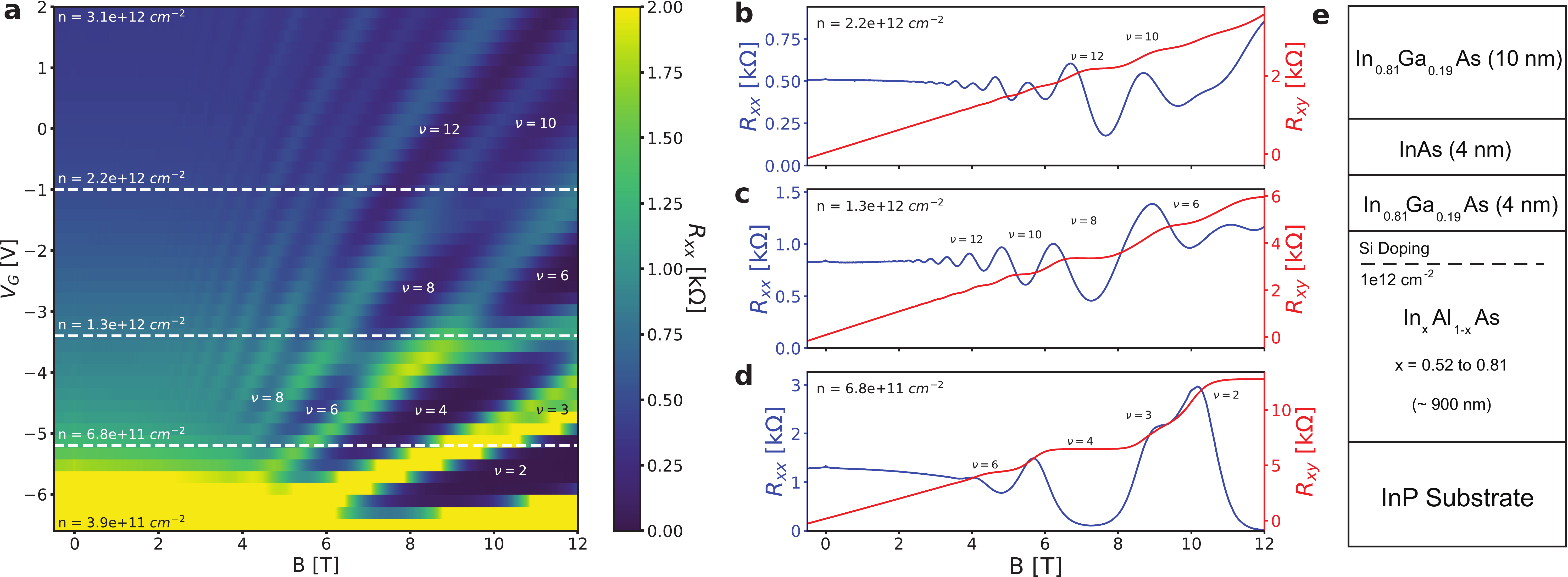}
\caption{(Color online) (a) Measured longitudinal resistance $R_{xx}$ vs magnetic field over a range of densities from $3.9 \times 10^{11} cm^{-2}$ to $3.1 \times 10^{12} cm^{-2}$.  Dashed lines indicate the traces that are shown in (b-d).  Integer quantum Hall states are labeled from complementary $R_{xy}$ data. (b-d) Longitudinal $R_{xx}$ and transverse $R_{xy}$ magnetotransport data at particular densities. The various integer quantum Hall states are labeled.  The left axis (blue trace) shows the longitudinal resistance $R_{xx}$ and the right axis (red trace) shows the transverse resistance $R_{xy}$.}
\label{fig:Rxx}
\end{figure*}

Two important material parameters of a 2DEG are the effective mass, $m^*$, and the effective $g$ factor, $g^*$.  These parameters dictate the response of a material to external electric and magnetic fields.  Their effect on device performance should be accounted for in the design of mesoscopic devices and realistic theoretical modeling.  Both $m^*$ and $g^*$ have been measured and calculated for bulk InAs \cite{InAsBulk} and for InAs QWs \cite{InAsQW,InAsQWSim}.  It is of particular interest that confinement of the electron wave function can strongly affect these values. Confinement becomes relevant when the 2DEG is placed near the surface, as is required for epitaxial contacts. In addition, narrow gap semiconductors can lead to strong non-parabolicity of the bands modifying the $m^*$ and $g^*$. However, to date, very few experimental studies have been performed to quantify the $m^*$ and $g^*$ in near surface InAs quantum wells. Here we report on these properties using Shubnikov-de Haas (SdH) oscillations and cyclotron resonance (CR) technique.

\subsection{II. Sample Growth and Preparation}
The samples were grown on a semi-insulating InP (100) substrate, using a modified Gen II molecular beam epitaxy system.  The In$_{x}$Al$_{1-x}$As buffer is grown at low temperature to help mitigate formation of dislocations originating from the lattice mismatch between the InP substrate and higher levels of the heterostructure \cite{Wallart05, ShabaniAPL2014, ShabaniMIT}. The indium content of In$_{x}$Al$_{1-x}$As is step-graded from $x =$ 0.52 to 0.81. Next, a delta-doped Si layer of $\sim 7.5 \times 10^{11}$ cm$^{-2}$ density is placed here followed by 6 nm of In$_{.81}$Al$_{.19}$As. The quantum well is grown next, consisting of a 4 nm thick layer of In$_{0.81}$Ga$_{0.19}$As layer, a 4 nm thick layer of InAs, and finally a 10 nm thick top layer of In$_{0.81}$Ga$_{0.19}$As.  A thin film of Al can be epitaxially grown on the final InGaAs layer. For the transport studies of the InAs quantum wells, Al films were selectively etched by Transene type-D solution while for optical studies Al was not grown from the beginning.  

\begin{figure*}[t]
\centering
\includegraphics[width=\textwidth]{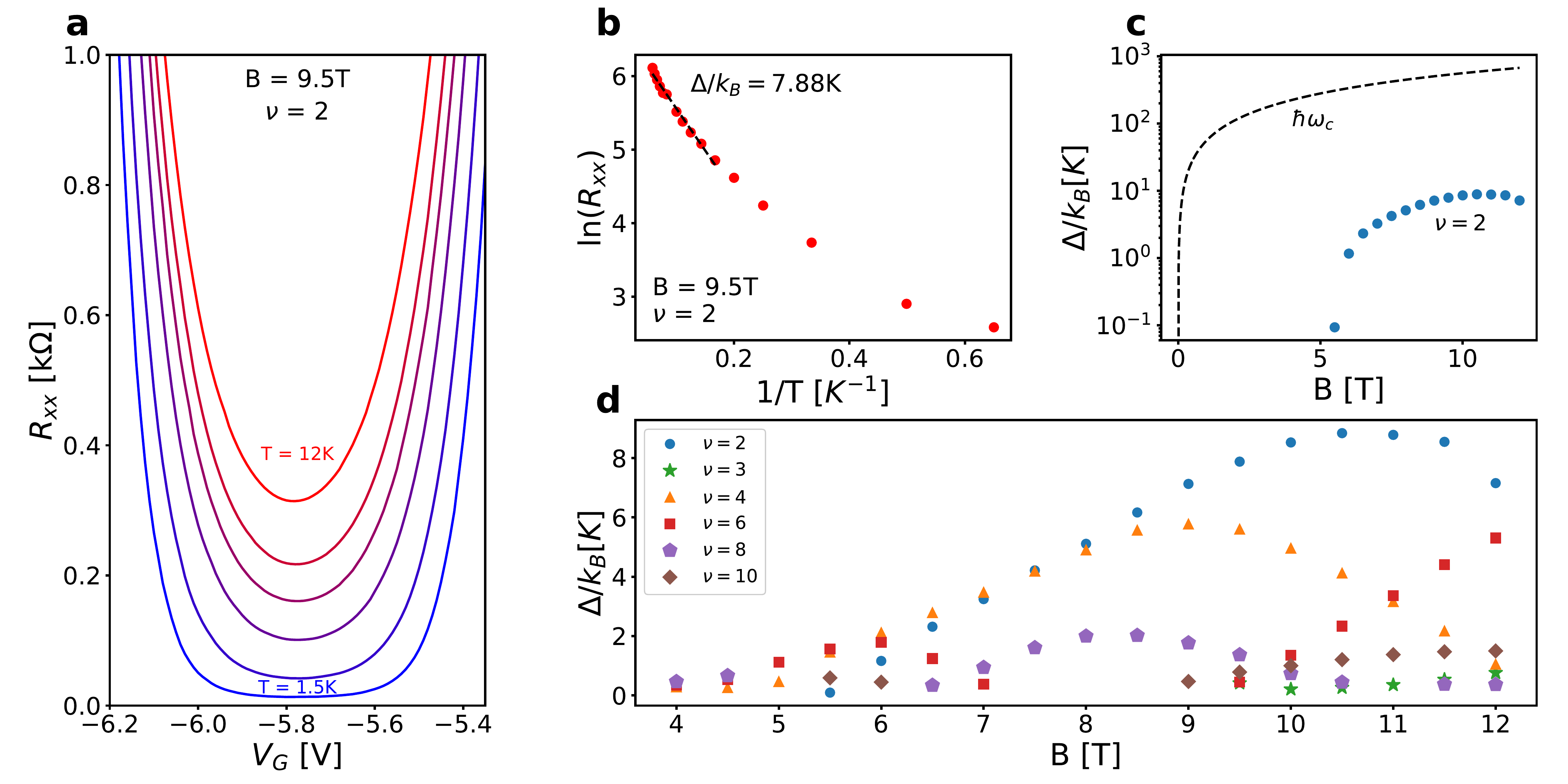}
\caption{(Color online) (a) Lifting of $\nu=2$ integer quantum Hall state longitudinal resistance as a function of gate voltage (density) at various temperatures between 1.5~K and 12~K.  (b) The natural logarithm of the minima in longitudinal resistance traces shown in (a). The higher temperature range data is linearly fitted and the gap is extracted from the slope.  (c) The gap energy shown on a logarithmic scale.  The $\nu=2$ gap is plotted for various magnetic fields.  This scale is used to highlight the large difference in expected range for the gap versus the measured gap. (d) The gap energy shown for various quantum hall states $\nu = {2,3,4,6,8,10}$.  The gaps are extracted in the manner exemplified in (a) and (b).  These gap energies when fit to the usual linear field dependance yield values for $m^*\sim0.2-2.1$ which are at least one order of magnitude higher than electron effective masses in general and landau level broadening of 10~K or less which does not represent the strong disorder expected from a two dimensional electron gas near the surface.}
\label{fig:TempGap}
\end{figure*}

\begin{figure*}[t]
\centering
\includegraphics[width=\textwidth]{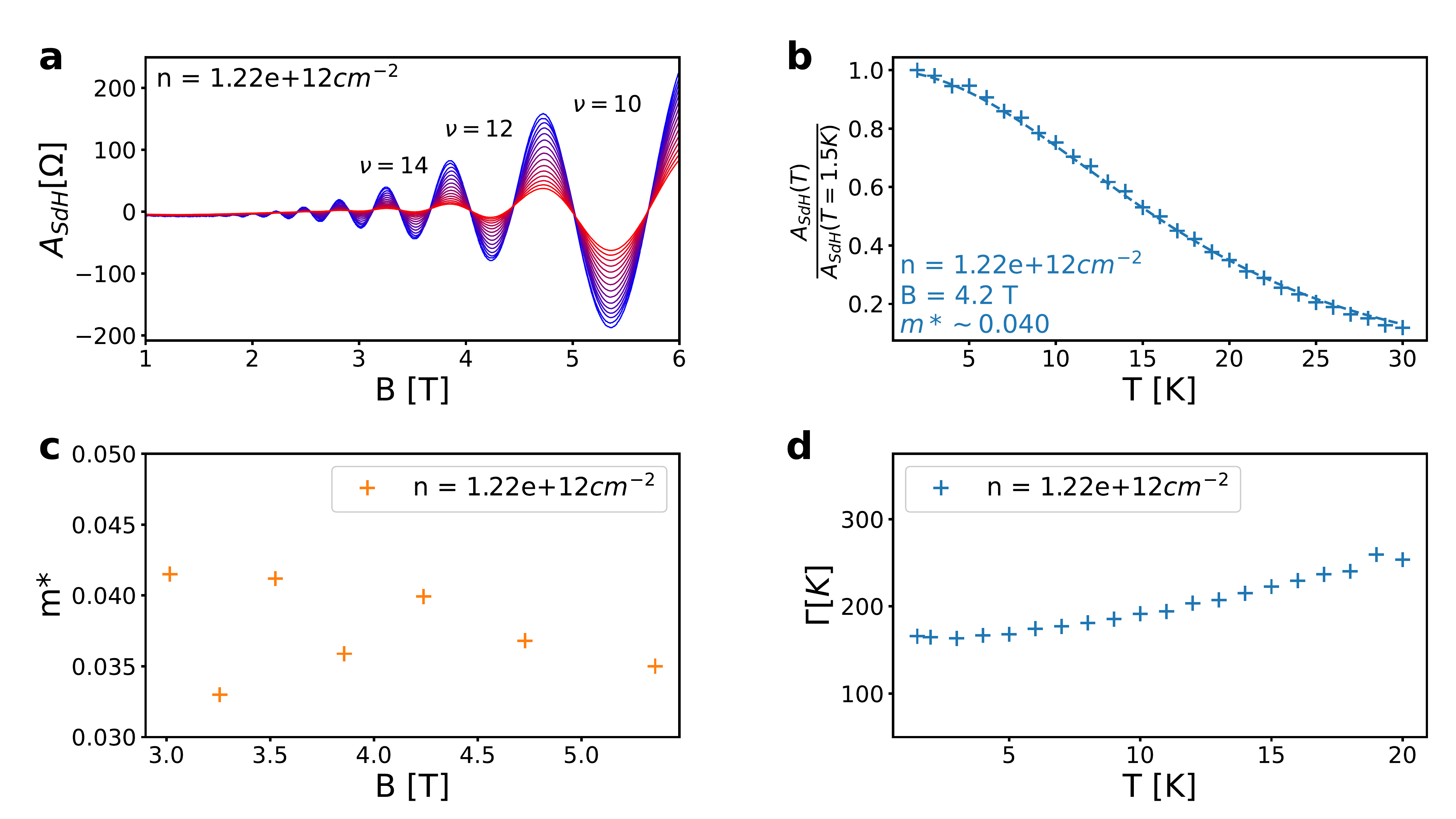}
\caption{(Color online) (a) The amplitude of Shubnikov-de Haas (SdH) oscillations obtained by subtracting the polynomial background from the longitudinal resistance.  Traces with largest amplitude (blue) were taken at temperature of 1.5~K and traces with lowest amplitude (red) were taken at a temperature of 30~K.  Traces of intermediate amplitude (and color) span the temperature range from 1.5 K to 30 K in steps of approximately 2~K.  Labeled quantum Hall states are extracted from Hall resistance.  (b) The normalized amplitude of SdH oscillations at B = 4.2~T. The points are data and the dashed line is the fit. The energy gap is extracted from the fits and used to calculate the effective-mass, m*. A value of m* = 0.04 is found for this oscillation extrema near B = 4.2 T.  (c)  $m^*$ values extracted from all reasonable oscillations.  (d) The Landau level broadening $\Gamma$ calculated from the quantum lifetime $\tau_q$ extracted for each temperature where an exponential envelope is fitted to the oscillations.}
\label{fig:SDH}
\end{figure*}

\subsection{III. Device Fabrication and Measurement Setup}
The samples used for our transport measurements were patterned using photolithography. The pattern used was an L-shaped Hall bar geometry allowing simultaneous measurement of longitudinal resistances ($R_{xx}$ and $R_{yy}$) and transverse resistance ($R_{xy}$). Chemical wet etching was performed after lithographic patterning leaving a 900~nm tall mesa. A 50~nm thick aluminum oxide (Al$_2$O$_3$) gate dielectric was then deposited on top of the Hall bar via atomic layer deposition. Gate electrodes were realized by subsequent deposition of 5~nm of titanium and 70~nm of gold. All measurements were performed inside a cryogen-free refrigerator with base temperature of 1.5~K with maximum magnetic field of 12~T.  Carrier densities are determined based on the slope of Hall data.

\subsection{IV. Measurement Results}
\subsection{A. Magnetotransport Measurements}
Figure~\ref{fig:Rxx}a shows the color-scale plot of longitudinal magnetotransport, $R_{xx}$, as a function of top gate voltage, $V_{G}$. The Landau level fan diagram is evident from the plot with  crossings observed at near $n = 1.3 \times 10^{12}$cm$^{-2}$ and 8~T and another near $n = 2.2 \times 10^{12}$cm$^{-2}$ and 12~T.  At lowest densities we only observe well developed integer quantum Hall states up to $n = 1.3 \times 10^{12}$cm$^{-2}$ ($V_{G}<$ -3~V). The first Landau level crossing appears near $V_{G}\sim$ -3~V where it signals occupation of the second electric subband. This is most evident as $\nu$ = 6 stays the same before and after the crossing in Fig.~\ref{fig:Rxx}a. Similar Landau level crossings have been studied extensively in GaAs 2DEGs \cite{Muraki2001,Gossard2006,Zhang2005,LiuPRB2011}. Three magnetotransport traces are shown in Fig. \ref{fig:Rxx}b-d. Longitudinal and Hall resistance as a function of magnetic field are plotted for $n = 2.2$, $1.3$, and $0.68 \times 10^{12}$ cm$^{-2}$. The beating in SdH oscillations clearly suggest occupation of two subbands at $n = 2.2 \times 10^{12}$cm$^{-2}$ where below the crossing clear quantum Hall states develop with vanishing longitudinal resistance at $n = 0.68 \times 10^{12} cm^{-2}$.

In a non-interacting quantum Hall system, the Landau level spacing increases with magnetic field as $\hbar\omega_c$ with $\omega_c=eB/(m^*m_e)$ where B is the magnetic field, and $m_e$ is the bare electron mass. Hence, measurements of energy gaps of integer quantum Hall states should be related to electron mass. Figure \ref{fig:TempGap}a shows the temperature dependence of longitudinal resistance as a function of gate voltage near the filling factor $\nu$ = 2 and at the magnetic field $B =$ 9.5~T. The natural logarithm of the minimum in resistance in a system with parabolic bands has a linear dependence on inverse temperature as shown in Fig. \ref{fig:TempGap}b \cite{lnR}. The energy gap is directly proportional to the magnitude of the slope.  We repeated these measurements as we varied the density and hence the position of $\nu$ = 2 in magnetic field. The results are shown in Fig.~\ref{fig:TempGap}c where extracted energy gaps are plotted as a function of magnetic field. For comparison, we also plot the energy gap expected from $\hbar\omega_c$ as a black dashed line. There is a large discrepancy between the measured and expected energy gap. If we allow electron mass to be a fitting parameter we obtain unrealistically high values of $m^* > 0.2$ for electrons.  We have also studied the energy gaps of filling factors $\nu$ = 3, 4, 6, 8, 10. Figure~\ref{fig:TempGap}d shows the energy gaps are between 0-10~K. All these values are much smaller than their corresponding $\hbar\omega_c$. The energy gaps for each filling factor first increase with magnetic field, then decrease, and eventually disappear near the Landau level crossings. For odd integer quantum Hall states, the Landau levels are split by the Zeeman energy $g^{*}\mu B$. Our data indicates that odd integers are mainly absent and only begin to develop at higher magnetic field ($\nu$ = 3 near 12 T) as shown in Fig.~\ref{fig:Rxx}a. Given the bulk g-factor in InAs (g = -14), the odd integers should have large enough energy gaps to be clearly observed. Their very weak presence is due to either modified $g^*$ or Landau level broadening due to disorder. To address this and the discrepancy of energy scales for gaps in even integer quantum Hall states we next measure the temperature dependence of the low magnetic field SdH oscillations where only free electrons contribute to the transport.

The SdH oscillation amplitude can be isolated by subtracting the background trend of the longitudinal resistance $R_{xx}$. Figure \ref{fig:SDH}a displays the amplitude of SdH, $A_{SdH}$ for a carrier density of $n = 1.22 \times 10^{12} cm^{-2}$.  Taking the points for a single minimum or maximum, normalized by our lowest temperature value, we can fit them to the formula $x/\text{sinh}(x)$ with $x =  2\pi^2 T/\Delta E$, where T is the temperature and $\Delta$E is the gap.  This allows us to calculate $m^* = \hbar e B / (m_e \Delta E)$.  Figure \ref{fig:SDH}b shows the data and fit for the oscillation near B~=~4.2~T from \ref{fig:SDH}a.  We have repeated these measurements for various filling factors to extract $m^*$ as shown in \ref{fig:SDH}c. The experimental values range between 0.035 - 0.05 with an average value near $m^*$ = 0.04. This is slightly higher than bulk values of our quantum well consisting of InAs and In$_{0.81}$Ga$_{0.19}$As with $m^*$ = 0.023 and 0.03 respectively. From the exponential envelope of the SdH oscillations we can also obtain the quantum lifetime and calculate the Landau level broadening, $\Gamma = \hbar/\tau_q$. Figure~\ref{fig:SDH}d shows $\Gamma$ for carrier density n = 1.22$ \times 10^{12}$ cm$^{-2}$. The Landau level broadening range is around 200~K for $n = 1.2 \times 10^{12}$ cm$^{-2}$. The broadening in the near surface InAs 2DEG is significantly larger than in buried InAs 2DEGs where $\Gamma$ is measured to be 5~K \cite{ShabaniMIT}. Here the surface scattering clearly dominates the other scattering mechanisms \cite{Kaushini2018}. Thankfully, the smaller electron mass in InAs enhances the energy scales and therefore enables us to resolve quantum Hall states. Our measured Landau level broadening could qualitatively describe the large discrepancy between energy gap measurements in the quantum Hall states and $\hbar\omega_c$. 

\begin{figure*}[htb]
\centering
\includegraphics[width=\textwidth]{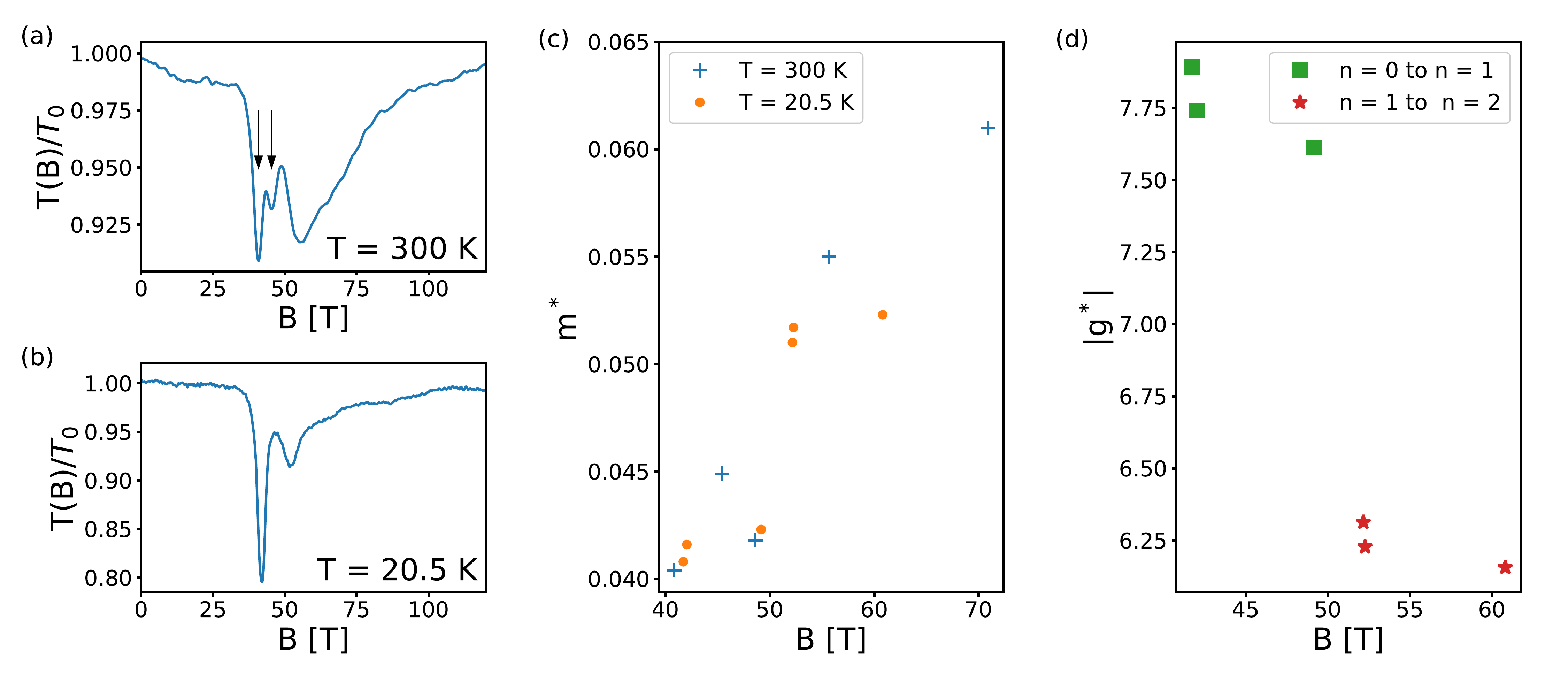}
\caption{(a) The normalized transmission of 10.6 $\mu$m excitation showing cyclotron resonance (CR) taken at T = 300~K (electron-active). The sample in this measurement has a density of $n=3.6 \times 10^{11}$ cm$^{-2}$.  The transitions indicated by arrows are attributed to the spin resolved CR transitions.  (b) The CR measurement displays a sharper transitions at T = 20.5~K. Unlike the measurements at 300 K, the spin resolved CR can not be resolved but the broader resonance at 55 T (the Landau level transition n = 1 to n = 2), observed at 300 K, shifts to lower fields and narrows down. 
(c) The effective mass $m^*$ as a function of magnetic field at T = 300~K and T = 20.5~K, demonstrate the non-parabolicity. (d) The absolute value of effective g-factor $g^*$ as a function of magnetic field at 20.5~K. 
}
\label{new-CR}
\end{figure*}

\subsection{B. Cyclotron Resonance Measurements}
A more direct way to measure $m^*$ is through infrared CR measurements using pulsed ultrahigh magnetic fields ($<$ 150 Tesla) generated by the single-turn coil technique \cite{Mag0,Mag1,Mag2}. The external pulsed magnetic field was applied along the growth direction and measured by a pick-up coil around the sample. The sample and the pick-up coil were placed inside a continuous flow helium cryostat. In this study, we employed infrared radiations from a CO$_2$ laser with wavelengths ranging from 9.2-10.6 $\mu$m.  The sample in this measurement has a density of $n=3.6 \times 10^{11}$ cm$^{-2}$.  The changes in transmission through the sample were collected using a fast liquid-nitrogen-cooled HgCdTe detector. A multi-channel digitizer placed in a shielded room recorded the signals from the detector and pick-up coil.

The spin resolved CR at 10.6 $\mu$m indicated by the two arrows in Fig. \ref{new-CR}a, separated by $\sim$ 4 Tesla, was observed at T = 300~K. This fact can be expected, as the Landau levels above the Fermi level can be occupied at T = 300~K, allowing the transitions between n = 0 and n = 1 for two different spins. In addition, in Fig. \ref{new-CR}a the broad resonance at $\sim$ 55 T represents a transition between n = 1 and n = 2 which is possible when the carrier lifetime allows time for a finite population of Landau level n = 1.  This transition is not predicted from the fixed Fermi energy, but can be attributed to the non-equilibrium electron distribution \cite{CR1,CR2}.

In Fig. \ref{new-CR}b, we present the CR measurements at 20.5~K with an excitation of 10.6~$\mu$m. The spin resolved CR was not observed indicating the states above the Fermi energy are no longer occupied. On the other hand, the broad resonance observed at $\sim$ 55~T and T = 300~K, which is due to the transition from n = 1 to n = 2, remained and narrowed.
Figure~\ref{new-CR}c summarizes our measurements for $m^*$ as a function of magnetic field at T = 300~K (crosses) and T = 20.5~K (filled circles). We note that although the single-turn coil is destroyed in each shot, the sample and pick-up coil remain intact, making it possible to carry out temperature and wavelength dependence measurements on the same sample. Figure~\ref{new-CR}c shows that the $m^*$ varied and increased monotonically with magnetic field. We measured $m^*$ = 0.04 near B = 40~T and $m^*$ = 0.061 near 70~T. Correspondingly we can estimate $g^*$ as a function of magnetic field using appropriate Landau level index using Eq.~1. In Fig.~\ref{new-CR}d we present absolute effective g-factor at 20.5 K as a function of magnetic field. 

\subsection{V. Landau Level Modeling}
\begin{figure}[htb]
\centering
\includegraphics [width = 0.45\textwidth] {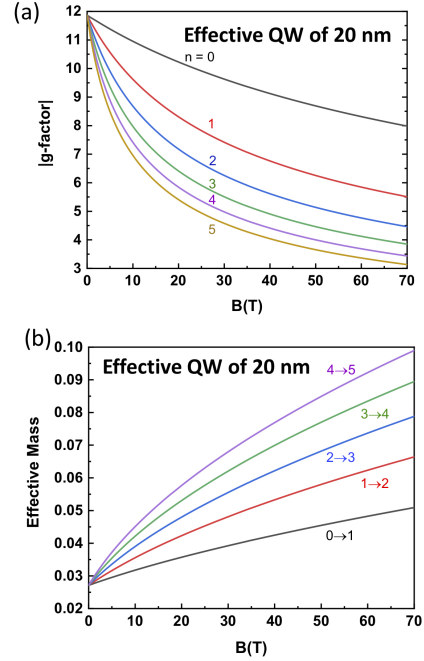}
\caption{ (a) Absolute values for the effective g-factor $g^*$ for the n = 0, ...5 Landau levels for the lowest subband for an InAs infinite square well with effective well width of 20 nm. One can see the sensitivity of the $g^*$ to the magnetic field and the Landau level index.  (b) The effective mass $m^*$(in units of the bare electron mass) for the n = 0, ...5 Landau levels for the lowest subband for an InAs infinite square well with 20 nm effective well width. Similar to $g^*$, $m^*$ varies as a function of the magnetic field and the Landau level index.}
\label {new-theory}
\end{figure}

Next we provide a simple theoretical model to understand $m^*$ and the Landau level fan diagram in InAs which has a non-parabolic conduction band.  Unlike the wide gap semiconductors such as GaAs, CR $m^*$ and $g^*$ may vary with subband index, Landau Level index, and external magnetic field. Beginning with expectations from the bulk and introducing confinement we can arrive at expressions for $m^*$ and $g^*$ (the details are presented in the Appendix):

\begin{equation}
\label{12}
	 {g^{*}_{j,n}} = \frac{{\left( {\varepsilon _{j,n}^ +  - \varepsilon _{j,n}^ - } \right)}}{{{\mu _B}B}}
\end{equation}

where $\varepsilon _{j,n}$ is the energy of the $n^{th}$ Landau level, for the $j^{th}$ subband index, and at magnetic field $B$. Plus and minus superscripts represent higher and lower Zeeman split energy bands respectively.  As shown in Fig. \ref{new-theory}a, $g^*$ depends on the subband index $j$, the Landau level $n$ as well as the magnetic field $B$. At zero magnetic field, the absolute value of $g^*$ = 12 is reduced from bulk value of $g^*$ = 14 due to confinement and monotonically decreases as magnetic field is increased. The rate depends on the Landau level index.

Similarly one can define $m^*$ obtained by CR as:
	
\begin{equation}
\label{13}	
	 m_{j,n}^{*, \pm } = \frac{{\hbar eB}/m_e}{{\left( {\varepsilon _{j,n + 1}^ \pm  - \varepsilon _{j,n}^ \pm } \right)}}
\end{equation}

We find that $m^*$, as shown in Fig. \ref{new-theory}b also depends on the $n^{th}$ Landau level, the $j^th$ subband index, and the magnetic field $B$ (we plot only the ($-$) solution for clarity). At zero magnetic field we see $m^{*}$ = 0.027 is larger than the bulk value of $m^*$ = 0.023 and increases monotonically as magnetic field is increased. These values are in close agreement with values derived from magnetotransport (over a small region 3~T to 5~T) and CR (40~T $<$ B $<$ 70~T).

\subsection{VI. Conclusion}
We have done magnetotransport and ultra high field cyclotron resonance characterization of surface InAs Quantum wells.  The density of these structures can be tuned and our magnetotransport measurement provides insight into the Landau level broadening and the quantum Hall energy gaps. By combining magnetotransport and cyclotron resonance measurements we can obtain conduction band effective mass $m^*$ at both low and high magnetic fields respectively. A band structure model which includes the effects of strong non-parabolicity and quantum confinement can describe the extracted $m^*$ from magnetotransport and cyclotron resonance measurements. We used our experimental CR $m^*$ values to determine the effective g-factor $g^*$ as a function of magnetic fields and Landau level index and these values are in a good agreement with the model presented here.

\textbf {Acknowledgment:}
The NYU team acknowledges partial support from U.S. Army Research Office agreements W911NF1810067 and W911NF1810115. and NSF Grants No. NSF-MRSEC 1420073 and No. NSF DMR - 1702594. G.A.K. and C.J.S. acknowledge support from the Air Force Office of Scientific Research under Award No. FA9550-17-1-0341. G.A.K. and B.A.M. acknowledge support from the Japanese visiting program of The Institute for Solid State Physics, The University of Tokyo. J.Y. acknowledges funding from the ARO/LPS QuaCGR fellowship reference W911NF1810067.

\section{Appendix: Simple model for electron mass and \MakeLowercase{g}-factor in a non-parabolic semiconductor.}

The derivation of the theoretical model accounting for non-parabolicity is described in this section.  In the absence of external magnetic field (and quantum confinement) a narrow gap semiconductor such as InAs has a conduction band energy, $\varepsilon$ vs. wavevector $k$ given by the dispersion relationship is given by:

\begin{equation}
\label{1}
	\varepsilon (1 + \alpha \varepsilon ) = \frac{{{\hbar ^2}{k^2}}}{{2m_o^*}} = \frac{{{\hbar ^2}\left( {k_x^2 + k_y^2 + k_z^2} \right)}}{{2m_o^*}}
\end{equation}

Here, $\alpha$ is the {\it non-parabolicity factor} given by

\begin{equation}
\label{2}
  \alpha  = 1/{\varepsilon _g}
\end{equation}
with $\varepsilon_g$ being the band-gap, and $m_o^*$ is the CR $m^*m_e$ {\it at the band edge} ($k=0$).  For small $\alpha\varepsilon$, the energy depends quadratically on $k$ while for large
$\alpha\varepsilon$, the energy depends linearly on $k$.

In the presence of a magnetic field in the $z$ direction, it can be shown \cite{mavroides1972magneto, lax1960cyclotron,  bowers1959magnetic} that one can write:

\begin{equation}
\label{3}
	\varepsilon (1 + \alpha \varepsilon ) = \frac{{{\hbar ^2}k_z^2}}{{2m_o^*}} + \left( {n + \frac{1}{2}} \right)\hbar {\omega _{c0}} \pm \frac{1}{2}{\mu _B}g_o^*B
\end{equation}

Here, $n$ is the Landau level index which can take on values [0,1,2,...].  $\omega_{c0}$ is the \textit{band-edge} CR frequency, given by:

\begin{equation}
\label{4}
	 {\omega _{c0}} = \frac{{eB}}{{m_o^*}}
\end{equation}

and
\begin{equation}
\label{5}
g_o^* = 2\left[ {1 + \,\,\left( {1 - \frac{1}{m^*}} \right)\,\,\frac{\Delta }{{3{\varepsilon _g} + 2\Delta }}} \right]
\end{equation}
is the {\it band-edge} $g^*$. $\Delta$ is the valence band spin-orbit splitting, and $\mu_B$ is the Bohr-magneton given by:

\begin{equation}
\label{6}
	 {\mu _B} = {\frac{{e\hbar }}{2m_e}}.
\end{equation}

Note that in the Bohr magneton, as opposed to the band-edge CR frequency, it is the bare electron mass that enters the expression.

To simplify, we set the RHS of Eq. \ref{3} to $K$

\begin{equation}
\begin {split}
\label{7}
 \varepsilon (1 + \alpha \varepsilon ) = \frac{{{\hbar ^2}k_z^2}}{{2m_o^*}} +\left( {n + \frac{1}{2}} \right)\hbar {\omega _{c0}}\\
  \pm \frac{1}{2}{\mu _B}g_o^*B = K
\end {split}
\end{equation}
and then solve for the energy $\varepsilon$.

\begin{equation}
\label{8}
 	\varepsilon  = \frac{{ - 1 \pm \sqrt {1 + 4\alpha K} }}{{2\alpha }}.
\end{equation}

The plus sign corresponds to the conduction band while the minus sign corresponds to the light hole in the valence bands. Quantum confinement will also affect both $g^*$ and $m^*$ for narrow gap materials.  To take into account quantum confinement, one quantizes $k_z$ as:

\begin{equation}
\label{9}
	 {k_z} = \frac{{2\pi }}{\lambda } = \frac{{j\pi }}{L}
\end{equation}

with $j$ a positive integer and $L$ the {\it effective} width of the quantum well.  Substituting into equation \ref{3} yields:

\begin{equation}
\begin {split}
\label{11}
 \varepsilon _{j,n}^ \pm (1 + \alpha \varepsilon _{j,n}^ \pm ) =\frac{{{\hbar ^2}{j^2}{\pi ^2}}}{{2m_o^*{L^2}}}
 +\left( {n + \frac{1}{2}} \right)\hbar {\omega _{c0}}\\
 \pm \frac{1}{2}{\mu _B}g_o^*B = K_{j,n}^ \pm
 \end {split}
\end{equation}

We assume an {\it effective} well width of 20 $nm$.  The gap at low temperatures is given by $\varepsilon_g = 0.4180$ while the spin orbit splitting is $\Delta = 0.38$ eV and the low temperature, band-edge effective mass is: $m_0^* = 0.023 m$.  From Eq. \ref{5}, we see this yields a band-edge $g_o^* = -14$.

\begin{figure}[htb]
\centering
\includegraphics[width = 0.5\textwidth]{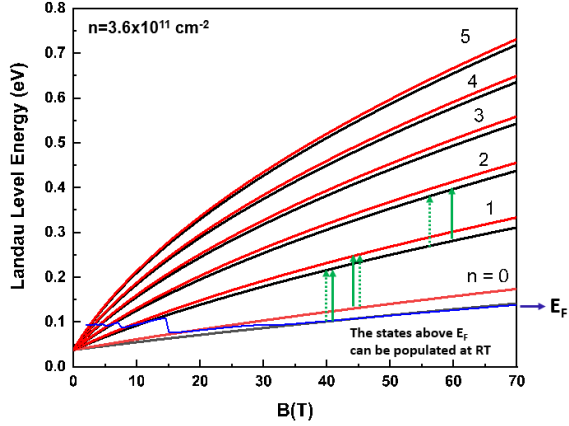}
\caption{
Calculated Landau Levels (n=0,...,5) for the lowest subband for a 20 nm InAs infinite square well in the simple model. 
}
\label {landau}
\end{figure}

The Landau fan energies in Eq.~\ref{11} can lead us to calculate and define $g^*$ for different Landau levels by:

\begin{equation}
\label{12}
	 {g_{j,n}} = \frac{{\left( {\varepsilon _{j,n}^ +  - \varepsilon _{j,n}^ - } \right)}}{{{\mu _B}B}}
\end{equation}

We can see that $g^*$ depends on the subband index $j$, the Landau level $n$ as well as the magnetic field $B$.

Similarly one can define $m^*$ by:
	
\begin{equation}
\label{13}	
	 m_{j,n}^{*, \pm } = \frac{{\hbar eB}/m_e}{{\left( {\varepsilon _{j,n + 1}^ \pm  - \varepsilon _{j,n}^ \pm } \right)}}
\end{equation}

Figure \ref{new-theory}(a,b) plots the $m^*$ and $g^*$ as a function of magnetic field and the Landau level index. We plot $m^*$ only for the lowest (-) solution.  Since $g^*$ will differ between Landau levels for a non-parabolic system.  The + and - effective masses will differ slightly and will lead to spin-split cyclotron resonance peaks under certain conditions.  The calculation shows that in presence of non-parabolicity both of these parameters depend on the subband index $j$, the Landau level $n$, and the magnetic field $B$. We note that assuming a smaller effective quantum well width (e.g. 12~nm) will shift $m^*$ to larger values (e.g. $\sim$ 0.035 at B = 0 T) and $g^*$ will shift smaller values ($\sim$ -9.5 at B = 0 T).

As shown in Fig. \ref{landau}, we have also calculated the Landau levels for the 1st subband. With the effective g-factor being negative, Red lines are spin down, Blacks are spin up.  The solid green arrows indicate the predicted CR transitions at 10.6 $\mu m$ and are in close agreement with experimental observations indicated by dashed green arrows.  While in the theory presented here, we considered the infinite potential well, the agreement between the theory and experiment is better at lower magnetic fields.  We should note that the Fermi level can be occupied at T = 300~K, allowing transitions between n = 0 and n = 1 for two different spins.  The spin resolved CR was not allowed at lower temperatures and the resonances above 50~T, in Fig.\ref{new-CR}a and Fig.\ref{new-CR}b are attributed to the transitions between n = 1 and n = 2 . These transitions are possible where the photo-excited carrier lifetime is long enough to populate the Landau level n = 1, even though the position of the Fermi level would not predict the transitions.

\begin{center}
{\bf References}
\end{center}

\bibliography{References_Shabani_Growth}
\pagebreak

\end{document}